\documentclass[12pt]{article}

\usepackage[margin=1in]{geometry}  
\usepackage{graphicx}              
\usepackage{amsmath}               
\usepackage{amsfonts}              
\usepackage{amsthm}                

\usepackage{lipsum}

\usepackage{caption}
\usepackage{subcaption}

\begin{document}
\date{}
\title{Creation of a tight PSF array for scanning structured illumination via phase retrieval}
\author{
  \small Alex Bardales$^1$, Qinggele Li$^{1,*}$, Bingzhao Zhu$^{1,2}$ Xiaodong Tao$^1$,\\
  \small Marc Reinig$^1$, Joel Kubby$^1$\\
\vspace{-3mm} 
  \scriptsize$^1$ \textit{W.M. Keck Center for Adaptive Optical Microscopy, Jack Baskin School of Engineering, University of California,}\\
\scriptsize \textit{Santa Cruz, California 95064, USA} \\
\vspace{-3mm} 
   \scriptsize $^2$ \textit{State Key laboratory of Modern Optical Instrumentation, College of Optical Science and Engineering and the Collaborative}\\ 
   \scriptsize \textit{ Innovation Center for Brain Science, Zhejiang University, Hangzhou, Zhejiang 310027, China} \\
   \scriptsize $^*$ qli34@ucsc.edu}

\maketitle 

\begin{abstract}
In this work, we propose a structured illumination (SI) method based on a two-photon excitation (TPE) scanning laser beam. Advantages of TPE methods include optical sectioning, low photo-toxicity, and robustness in the face of sample induced scattering. We designed a novel multi-spot point spread function (PSF) for a fast, two-photon scanning SIM microscope. Our multi-spot PSF is generated with a phase retrieval algorithm. We show how to obtain the phase distribution and then simulate the effect of this distribution on a spatial light modulator (SLM), which produces the multi-spot PSF in the object plane of the microscope. We produce simulations that show the viability of this method. The results are simulated and a multi-spot PSF scanning SIM microscope is proposed.
\end{abstract}

\section{Introduction}

Many important features in biological microscopy are on the order of 10's to 100's of nanometers~\cite{huang2010breaking}. To image these tiny features, scanning electron microscopy can be used, but this is undesirable since that class of microscopes has strict requirements on sample preparation. The primary disadvantage, from a biological point of view, is that the samples are fixed. On the other hand, if a light microscope is used properly, biologists obtain \textit{in vivo} imaging. When compared to electron microscopes, the main disadvantage of the light microscope is its resolving power. The wave nature of light fundamentally limits a microscope's resolution, which is around half the wavelength of excitation light~\cite{born2000principles}. To overcome this limit, super-resolution imaging (SR) has been implemented. Another problem for biological imaging is sample-induced aberration. The way to overcome this barrier is through use of adaptive optics (AO).

In the past two decades, there have been many SR microscopes invented, and this field is growing with each year. Stimulated emission depletion microscopy (STED) was invented in 1994 and works by exciting a small fluorescent area in a sample and then depleting a tight area around the centroid of excitation, confining emission light to a very small region~\cite{hell1994breaking}. Stochastic methods include photo-activated localization microscopy (PALM) and stochastic optical reconstruction microscopy (STORM) that operate under the principle of statistically locating fluorescent markers that are photo-switchable or photo-activatable~\cite{betzig2006imaging,rust2006sub}.

Structured illumination microscopy (SIM) is yet another SR technique~\cite{gustafsson2000surpassing}. Linear SIM only gives a modest super-resolution factor of 2~\cite{schermelleh2010guide}, but is less invasive and faster than the other SR microscopes. SIM works by projecting illumination patterns onto the sample such that high spatial frequency information which is beyond the microscope's resolving power is aliased into the pass-band of the microscope. A convenient illumination pattern is therefore sinusoidal, because this has a simple Fourier domain representation. SIM is typically applied to widefield imaging, 
so image quality is often degraded by out of focus light, aberration, and scattering. This makes imaging a densely labeled, thick sample all but impossible.


Two-photon excitation (TPE) microscopy rejects out of focus light inherently~\cite{denk1990two}. Recently, a new SIM-based method of microscopy that incorporates TPE has been implemented~\cite{schermelleh2010guide}. In this method, called scanning SIM, illumination patterns are scanned onto the object plane. In that work, the authors use a TPE method with an intensity modulation that emulates a sinusoidal pattern in the object plane. Emission light from the sample is then collected through both a photo-multiplier tube (PMT) and a CCD array. An image is built up by summing individual contributions from the excitation light. The price to pay when using this technique is that it is slow when compared to conventional SIM. In conventional SIM, a typical setup requires 9 widefield images for a single reconstruction. If an image is to be built up into $N\times N$ pixels, then the number of PMT readings is increased by a factor of $N^2$. If one were to attempt to reduce the image acquisition time for scanning SIM, the $N^2$ factor would be a good place to start. Faster scanning SIM utilizes widefield imaging with a CCD array. TPE scanning SIM has been implemented in~\cite{york2012resolution}, with widefield image acquisition. Later on, the same group, went on to build a TPE scanning SIM microscope which used a pinhole array coupled with a microlens array~\cite{york2013instant}, a setup which allowed several excitation spots to be imaged at once. 

Scanning SIM is attractive because the PSF of the microscope is contained within a small region in the object plane, which opens up the technique to adopting AO. In this study, we propose an alternative way to build up the scanning SIM illumination pattern that is both fast and suitable for AO. With the phase retrieval algorithm given in~\cite{gerchberg1972practical}, we can create a multi-spot PSF in the object plane by using a spatial light modulator (SLM). 
We propose a TPE microscope since it rejects out of focus light, reduces photo bleaching and photo-toxic effects, and is good for optical sectioning. We also propose a multi-spot PSF in order to speed up the imaging process. We simulate an illumination pattern that is suitable for a SIM reconstruction with TPE. Furthermore, since the multi-spot PSF is contained within a small region in the object plane, we can use adaptive optics to make local wavefront corrections within the sample. This will allow SIM to be used in deep tissue imaging.

The rest of the paper is organized as follows. In Section~\ref{meth}, we discuss the mathematical framework of SIM, and also method of obtaining our simulations. In Section~\ref{res}, we discuss the results of our simulations, as well as the implications for using this method for scanning SIM. We conclude with Section~\ref{concl}, with a summary of the present work and discussion of future work.

\section{Methods}\label{meth}
\subsection{SIM framework}
We begin with the framework for SIM, which follows, roughly, the work done in~\cite{gustafsson2000surpassing,shroff2008otf,gustafsson2008three}. In this discussion, we will refer to three different images. The first is the image that which we would like to see, but are not able to, because of the low-pass filtering action of microscope objective lens. We will refer to this image as the ground truth, the sample under observation which contains unobservable, high frequency information. The image that we can observe with the microscope is called just that: the observable image. The third image is called the illumination pattern. This is the structured pattern that will be used to heterodyne the high frequency information into the pass band of the microscope system. The Fourier transform (FT) of the PSF is known as the optical transfer function (OTF).

For any point in real space, $\mathbf{r}$, we denote the ground truth by $D(\mathbf{r})$ and the observable image by $D'(\mathbf{r})$. These functions are related to the OTF, $H(\mathbf{r})$ by the following (for brevity we have combined this equation with its Fourier dual):

\begin{equation}\label{conv1}
D'(\mathbf{r}) = D(\mathbf{r} )\otimes H(\mathbf{r})  \Longleftrightarrow D'(\mathbf{k}) = D(\mathbf{k})\cdot H(\mathbf{k}),
\end{equation}
where $\otimes$ denotes the convolution operator, and $\mathbf{k}$ being a point in reciprocal space.

Next, we address the illumination pattern, $I(\mathbf{r})$. If we make this illumination pattern a sinusoid, then its illumination intensity can be given by

\begin{equation}\label{illumination}
I(\mathbf{r}) = \frac{1}{2}[1 + \cos(2\pi \mathbf{p} \cdot \mathbf{r} + \phi_n)],
\end{equation}
where $\mathbf{p}$ is the spatial frequency vector of the illumination pattern, and $\phi_n$ is a phase shift. For conventional SIM reconstructions, $n$ ranges from 1 to 3.

Another convenience of this illumination pattern is seen in its spatial-frequency domain representation which is a sum of three of delta functions:

\begin{equation}
I(\mathbf{k}) = \frac{1}{2}\delta (\mathbf{k}) + \frac{1}{4}\delta(\mathbf{k}-\mathbf{p}) + \frac{1}{4}\delta(\mathbf{k}+\mathbf{p})],
\end{equation}

and if we mix (or multiply) Equation \ref{illumination} with the ground truth image, then we obtain

\begin{equation}\label{conv2}
D'(\mathbf{r}) = D(\mathbf{r})I(\mathbf{r}) \otimes H(\mathbf{r}) \Longleftrightarrow D'(\mathbf{k}) = [D(\mathbf{k}) \otimes I(\mathbf{k})] H(\mathbf{k}).
\end{equation}

The convolution theorem allows us to write the spectrum of $D'(\mathbf{k})$ as

\begin{equation}\label{conv2_k}
D'(\mathbf{k}) = \frac{1}{2}\left[D(\mathbf{k}) + \frac{1}{2}D(\mathbf{k}-\mathbf{p}) e^{-j\phi} + \frac{1}{2}D(\mathbf{k}+\mathbf{p}) e^{+j\phi}\right]\cdot H(\mathbf{k}),
\end{equation}
so we see that under the illumination pattern $I(\mathbf{r})$, the spectrum of the observable image is a linear, phase-shifted combination of the ground truth. Here we note that lower spatial frequencies of $D(\mathbf{k})$ already fall within the pass band of the OTF, $H(\mathbf{k})$, but $D(\mathbf{k}\pm\mathbf{p})$ do not lie within the pass band. 

If we use different phases, then we can have a series of measurements that can be represented in matrix form as

\begin{equation}\label{matrix}
\left[ \begin{array}{c} D_1'(\mathbf{k}) \\ D_2'(\mathbf{k}) \\ \vdots \\ D_N'(\mathbf{k}) \end{array} \right] =  \begin{bmatrix} 1 & e^{j\phi_1} & e^{-j\phi_1} \\ 1 & e^{j\phi_2} & e^{-j\phi_2} \\ \vdots & \vdots & \vdots \\1 & e^{j\phi_N} & e^{-j\phi_N}  \end{bmatrix}  \times \left[\arraycolsep=1.4pt\def\arraystretch{1.6} \begin{array}{c} \frac{1}{2}H(\mathbf{k})D(\mathbf{k}) \\ \frac{1}{4}H(\mathbf{k})D(\mathbf{k}-\mathbf{p}) \\  \frac{1}{4}H(\mathbf{k})D(\mathbf{k}+\mathbf{p})  \end{array} \right].
\end{equation}
However, in this work we will use only three phases, 0, $2\pi/3$, and $4\pi/3$, as this is the minimum number of known variables needed to solve for three unknown equations.

\sloppy
It is the second two rows of the rightmost matrix in Equation~\ref{matrix}, the terms with~$1/4H(\mathbf{k})D(\mathbf{k}\pm\mathbf{p})$, that contain the super-resolution information. By heterodyning these frequencies into the passband of the microscope, the SR information can be recovered. In the reconstruction step, we move these SR spatial frequency components, $D(\mathbf{k} \pm \mathbf{p})$, back to their original location in reciprocal space. Since this shift operation involves overlapping regions, there may be more than one estimate for a given point in reciprocal space. A weighted average can be implemented to obtain an overall Fourier representation of the SR image. A Wiener filter is also typically used to suppress numerical instability, and artifacts may appear. In this work, we do not consider additive noise, because this process is well understood and incorporated into SIM reconstruction algorithms.~\cite{shroff2008otf,lal2016structured}.

This mathematical framework demonstrates how SIM effectively increases the pass-band of the OTF. After recovering the enlarged Fourier domain representation of the object, we perform an inverse Fourier transform (IFT). In this way, we can recover the high-spatial frequency information that was originally outside the pass-band of the microscope OTF. For more information about the SIM framework, the reader is directed to~\cite{gustafsson2000surpassing,gustafsson2008three}.

\subsection{Simulation framework}

\begin{figure}[t]
    \centering
    \begin{subfigure}{0.3\textwidth}
        \includegraphics[width=\textwidth]{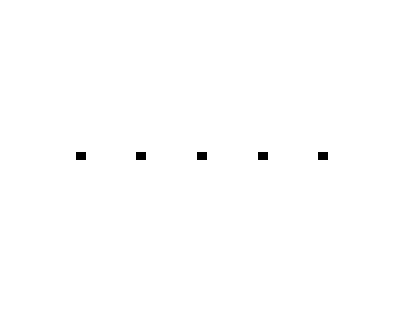}
        \caption{Target PSF 1}
        \label{fig:tpsf2}
    \end{subfigure}
    ~ 
    \begin{subfigure}{0.3\textwidth}
        \includegraphics[width=\textwidth]{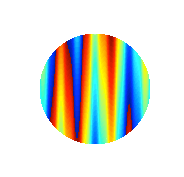}
        \caption{Phase solution}
        \label{fig:pp2}
    \end{subfigure}
    ~ 
    \begin{subfigure}{0.35\textwidth}
        \includegraphics[width=\textwidth]{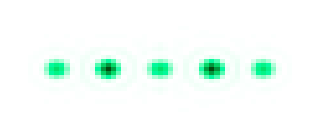}
        \caption{PSF in object plane}
        \label{fig:ppsf2}
    \end{subfigure}
    \caption{Non-ideal proposal PSF for multi-spot scanning SIM. The five dots in the target PSF are close enough that coherent interference distorts the intensity of each spot in the object plane.}\label{fig:bad_prop}
\end{figure}

\begin{figure}[t]
    \centering
    \begin{subfigure}{0.3\textwidth}
        \includegraphics[width=\textwidth]{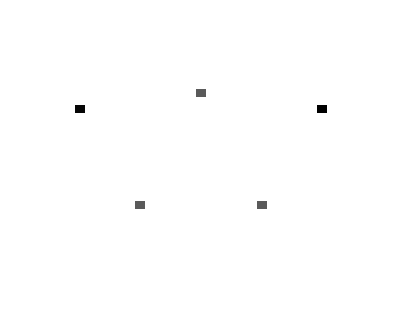}
        \caption{Target PSF 2}
        \label{fig:tpsf1}
    \end{subfigure}
    ~ 
    \begin{subfigure}{0.3\textwidth}
        \includegraphics[width=\textwidth]{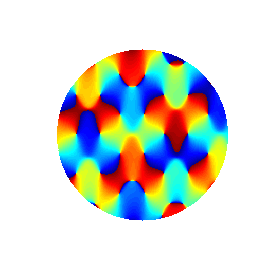}
        \caption{Phase solution}
        \label{fig:pp1}
    \end{subfigure}
    ~ 
    \begin{subfigure}{0.3\textwidth}
        \includegraphics[width=\textwidth]{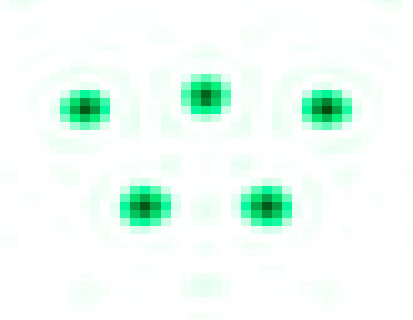}
        \caption{PSF in object plane}
        \label{fig:ppsf1}
    \end{subfigure}
    \caption{A better proposal PSF for multi-spot scanning SIM. By using an offset in the vertical direction, coherence is no longer a problem in the object plane. The offset is not an issue after scanning this PSF through the object plane.}\label{fig:good_prop}
\end{figure}

Conventional SIM is usually a widefield microscopy technique. The robustness of the reconstruction depends on the fidelity of the illumination patterns (cf Equation~\ref{illumination}). That is to say, any deviation from the ideal sinusoidal illumination pattern, the spectral leakage in the frequency domain will degrade the reconstruction, since Equation~\ref{conv2_k} will be violated. Therefore, a clean, high-contrast illumination pattern is essential for a SIM reconstruction . Another source of error in SIM reconstruction is an incorrect estimation of phase and spatial frequency of the illumination patterns. In this work, spatial frequency we estimate using Fourier domain techniques. Phase can be estimated using an autocorrelation maximization routine, as in~\cite{lal2016structured}.

Pattern projection onto the object can be challenging in and of itself, but when imaging deep into a sample, sample-induced aberrations blur the illumination pattern, making SIM reconstruction difficult. Use of Adaptive Optics (AO) can help, but wavefront correction is limited to a small area known as the isoplanatic patch, which typically does not cover the entire field of view~\cite{kubby2013adaptive}. For a typical image of a drosophila embryo, the isoplanatic patch has been found to be around 20$\mu$m~\cite{tao2011adaptive}. Therefore, we aim to keep our PSF contained within a region of this size in the object plane.

In two-photon excitation (TPE) microscopy, the excitation light is focused into a small volume in the object plane, similar to a confocal laser scanning microscope. The theoretical PSF of a single photon confocal system is modeled by a Bessel function of the first kind~\cite{gu1996principles}. The TPE system is nonlinear, and its PSF is proportional to the square of the intensity of the single photon PSF~\cite{sheppard1990image}. Therefore, our multi-spot PSF must be squared in our simulations. 

\begin{figure}
\begin{center}
\includegraphics[scale=0.5]{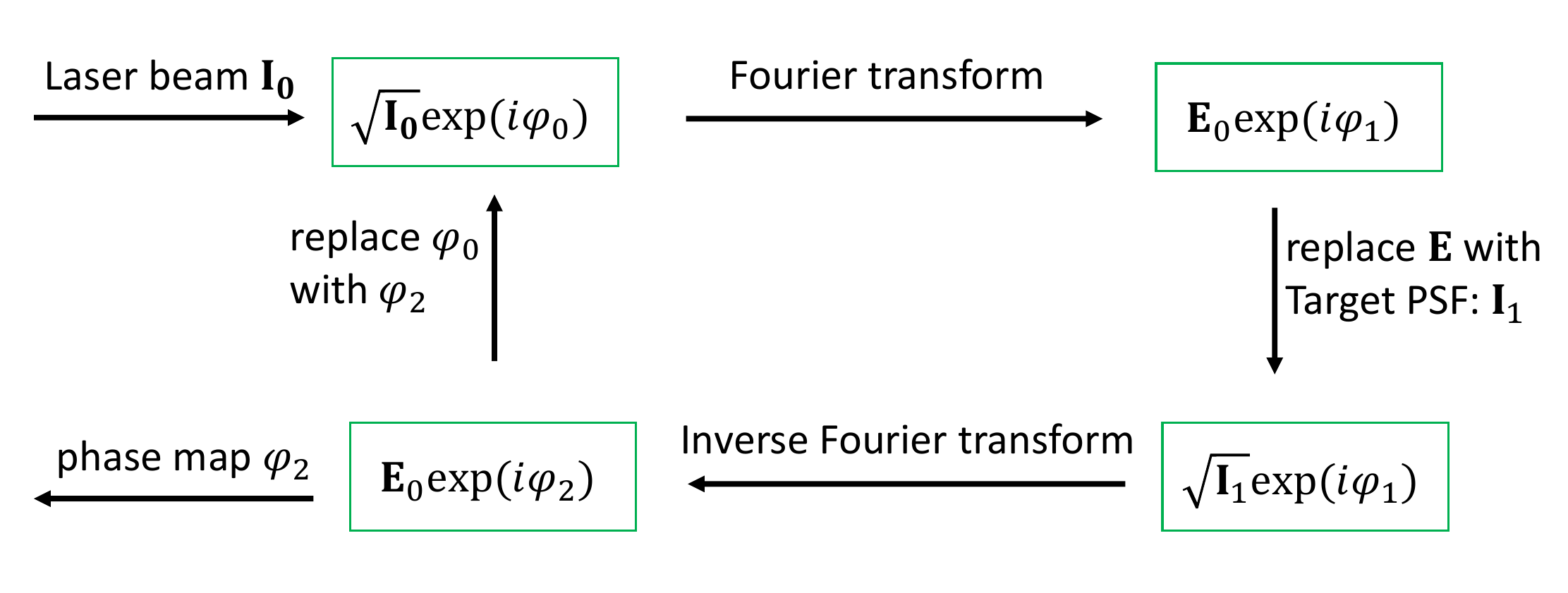}
\caption{Block diagram of the iteration step of Gerchberg-Saxton algorithm for phase retrieval.} 
\label{fig:algo}
\end{center}
\end{figure}

We use the Gerchberg-Saxton phase retrieval algorithm from~\cite{gerchberg1972practical}, implemented in Matlab. We begin by proposing a desired PSF that meets our requirements for scanning SIM. Next we perform the phase retrieval algorithm to derive a phase distribution for the SLM. We then generate a PSF from the phase distribution in order to emulate the scanning operation in software. Figure~\ref{fig:algo} shows a block diagram of the iterative loop. In a typical SIM reconstruction, here are three phases of illuminations patterns, and at least two rotations. With our method, the illumination pattern can be phase-shifted by changing the scanning path with galvo mirrors. Likewise, to rotate the illuminate pattern, the SLM distribution can be rotated, and the beam steered accordingly.

\section{Results and discussion}\label{res}

To illustrate the simulation process, we propose a multi-spot PSF as shown in Figure~\ref{fig:bad_prop}. The proposed PSF is simply an array of five points arranged in a linear fashion. In making this proposed PSF, the hope was that the derived phase map would produce a set of points that have an even intensity distribution. In Figure~\ref{fig:ppsf2} we can see that this is not the case. If we were to use this generated PSF for a scanning SIM reconstruction, the illumination pattern would be roughly sinusoidal, but with additional frequency components that make the pattern unsuitable for a SIM reconstruction. The uneven intensity distribution among the five spots is likely a result of coherent interference of the laser source. Thus, a better proposed PSF should increase the distance between the points so that coherent interference is decreased.

\begin{figure}
\begin{center}
\includegraphics[scale=0.5]{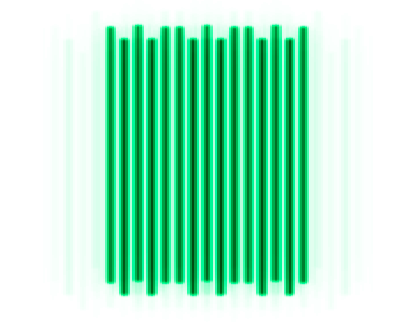}
\caption{Illustration of scanning SIM using a multi-spot PSF. The PSF here is a 2D scan of Figure \ref{fig:ppsf1}, obtained from the phase pattern shown in Figure \ref{fig:pp1}.}
\label{fig:buildup}
\end{center}
\end{figure}

In order to create a more even distribution among the PSF spots, some spacing should be added in between the pixels in the proposed PSF. This is what we see in Figure~\ref{fig:tpsf1}. We add some offset in the vertical direction in order to mitigate the effect of coherent interference. We can also weight the different points in the proposed PSF so that the intensity distribution of the generated PSF is more uniform. The offset in the vertical direction of the proposed PSF introduces an asymmetry, but this will not be an issue due to the scanning operation. In fact, we can illustrate this point by simulating this very operation. In Figure~\ref{fig:buildup}, we can see a small sample of the effective illumination pattern by scanning the generated PSF from Figure~\ref{fig:ppsf1}. It is apparent that this illumination pattern is very close to sinusoidal, although there is a ghosting effect due to the presence of side lobes. In practice, this effect will not be present since the illumination pattern can be made arbitrarily large by the scanning operation. We can scan an area that is larger than the region of interest, and then crop the SIM images before performing the SIM reconstruction algorithm.

\begin{figure}
\begin{center}
\includegraphics[scale=0.5]{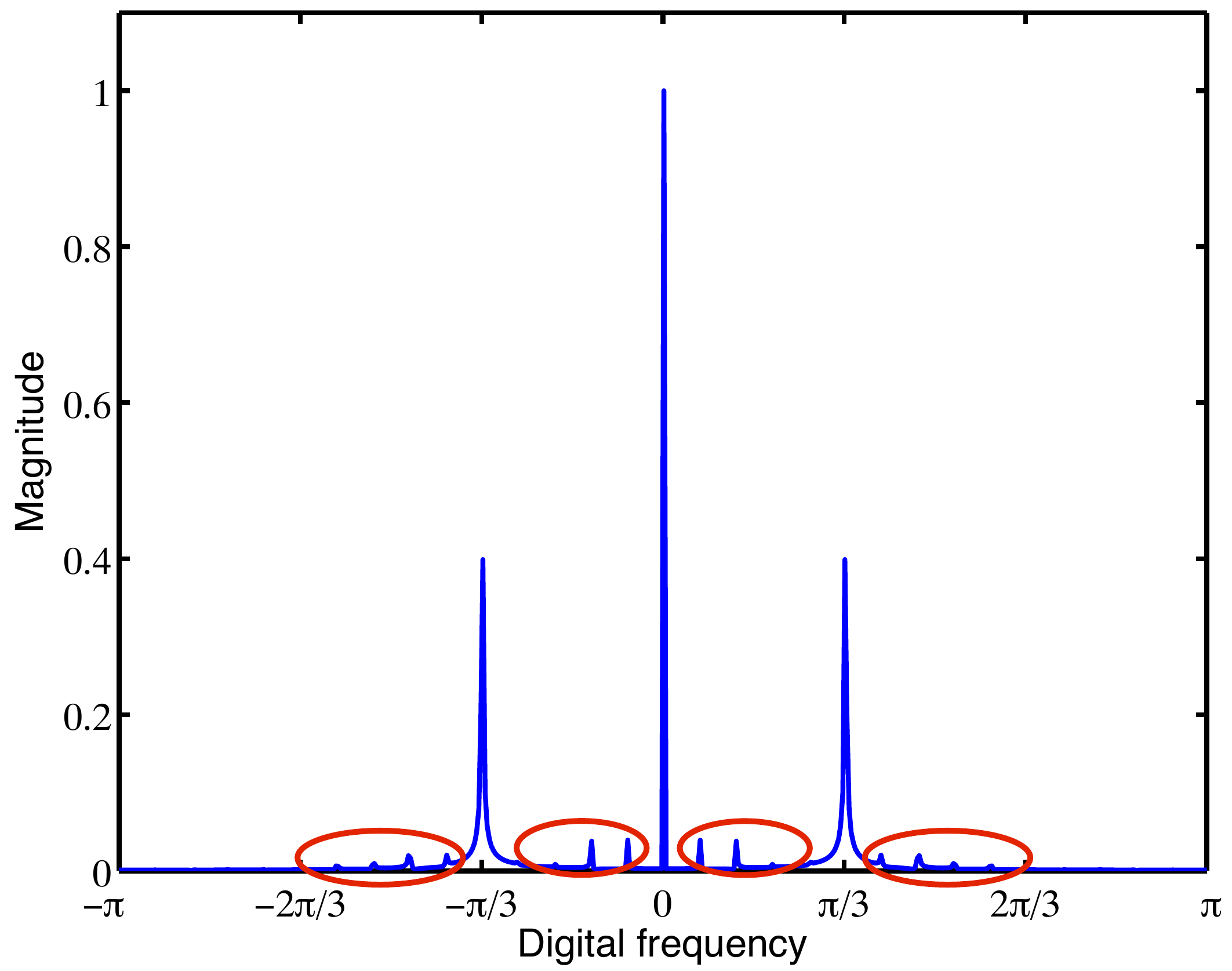}
\caption{The Fourier domain representation of the illumination pattern obtained by the multi-spot scanning SIM method. The conventional SIM illumination patterns would only contain the peaks at frequencies $\pm\pi/3$, and 0. Here, the spectral leakage is noted by the red ovals. In principal, the phase-retrieval algorithm could produce a multi-spot PSF whose sinusoidal scanning pattern is arbitrarily close to the ideal sinusoid.}
\label{fig:fft}
\end{center}
\end{figure}

\begin{figure}
    \centering
    \begin{subfigure}{0.3\textwidth}
        \includegraphics[width=\textwidth]{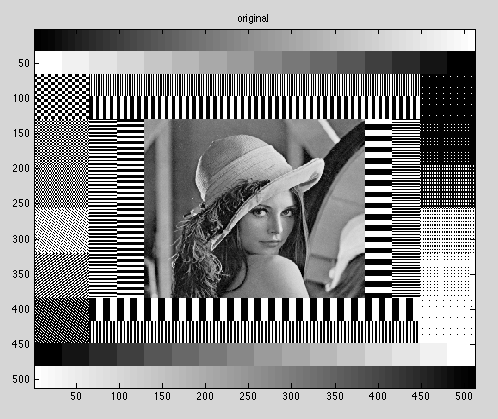}
        \caption{Ground truth image}
        \label{fig:orig}
    \end{subfigure}
    ~ 
    \begin{subfigure}{0.3\textwidth}
        \includegraphics[width=\textwidth]{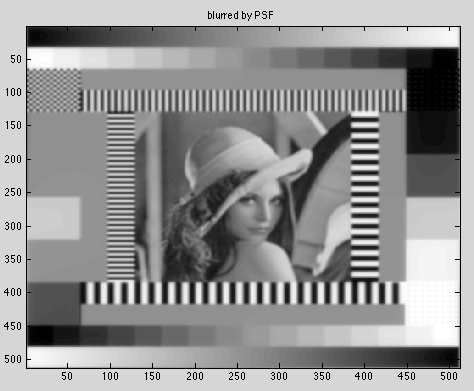}
        \caption{Blurred image}
        \label{fig:blur}
    \end{subfigure}
    ~ 
    \begin{subfigure}{0.3\textwidth}
        \includegraphics[width=\textwidth]{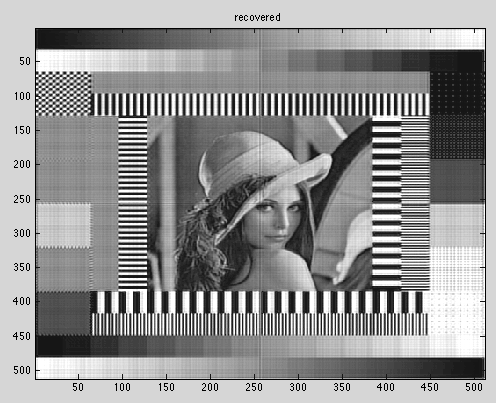}
        \caption{Reconstruction}
        \label{fig:recon}
    \end{subfigure}
    \caption{A demonstration of the reconstruction with our method. Image (a) is the original image, from~\cite{lal2016structured}. Image (b) is what would be visible with a conventional, widefield microscope. The SIM reconstruction (c) is shown using the pattern obtained in this section.}\label{fig:rec}
\end{figure}

In Figure~\ref{fig:fft}, we show the 1D Fourier representation of our 2D sinusoidal pattern. If our generated pattern were truly sinusoidal, then only the peaks present at $\pm\pi/3$ would be present. However, we note that the areas denoted by the red ovals have some noise. In practice, we would continue to refine the multi-spot PSF using the phase retrieval algorithm to come up with a better PSF. We illustrate a SIM reconstruction, using the pattern shown in Figure~\ref{fig:buildup} as a basis. In Figure~\ref{fig:rec}, the test image is blurred with the conventional PSF of the microscope system (\ref{fig:blur}). Equivalently, the image is filtered in the Fourier domain by the optical transfer function of the microscope. Finally, the SIM reconstruction is shown (\ref{fig:recon}). 

\section{Conclusion}\label{concl}

In this work we simulated a scanning SIM reconstruction method based on PSF shaping in the object plane. The PSF was constructed with a phase retrieval algorithm, which in turn generated a real PSF in the object plane. This generated PSF can be written into in the object plane of a microscope with a spatial light modulator. The PSF is large enough to offer an increase in the speed of scanning SIM, yet small enough to be fully contained within an isoplanatic patch in an AO system. We plan on carrying out this work in a TPE microscope with adaptive optics for obtaining SIM reconstructions in thick, densely labeled samples, such as in a drosophila embryo. 

\section{Acknowledgments}

The results presented herein were obtained at the W.M. Keck Center for Adaptive Optical Microscopy (CfAOM) at University of California Santa Cruz. The CfAOM was made possible by the generous financial support of the W.M. Keck Foundation. This material is also based upon work supported by the National Science Foundation under Grant. Numbers 1353461 \& 1429810. Any opinions, findings, and conclusions or recommendations expressed in this material are those of the authors and do not necessarily reflect the views of the National Science Foundation.

\pagebreak
\bibliographystyle{unsrt}
\bibliography{simbib} 

\begin{thebibliography}{10}

\bibitem{huang2010breaking}
Bo~Huang, Hazen Babcock, and Xiaowei Zhuang.
\newblock Breaking the diffraction barrier: super-resolution imaging of cells.
\newblock {\em Cell}, 143(7):1047--1058, 2010.

\bibitem{born2000principles}
Max Born and Emil Wolf.
\newblock {\em Principles of optics: electromagnetic theory of propagation,
  interference and diffraction of light}.
\newblock CUP Archive, 2000.

\bibitem{hell1994breaking}
Stefan~W. Hell and Jan Wichmann.
\newblock Breaking the diffraction resolution limit by stimulated emission:
  stimulated-emission-depletion fluorescence microscopy.
\newblock {\em Optics letters}, 19(11):780--782, 1994.

\bibitem{betzig2006imaging}
Eric Betzig, George~H. Patterson, Rachid Sougrat, O.~Wolf Lindwasser, Scott
  Olenych, Juan~S. Bonifacino, Michael~W. Davidson, Jennifer
  Lippincott-Schwartz, and Harald~F. Hess.
\newblock Imaging intracellular fluorescent proteins at nanometer resolution.
\newblock {\em Science}, 313(5793):1642--1645, 2006.

\bibitem{rust2006sub}
Michael~J. Rust, Mark Bates, and Xiaowei Zhuang.
\newblock Sub-diffraction-limit imaging by stochastic optical reconstruction
  microscopy (storm).
\newblock {\em Nature methods}, 3(10):793--796, 2006.

\bibitem{gustafsson2000surpassing}
Mats~G.L. Gustafsson.
\newblock Surpassing the lateral resolution limit by a factor of two using
  structured illumination microscopy.
\newblock {\em Journal of microscopy}, 198(2):82--87, 2000.

\bibitem{schermelleh2010guide}
Lothar Schermelleh, Rainer Heintzmann, and Heinrich Leonhardt.
\newblock A guide to super-resolution fluorescence microscopy.
\newblock {\em The Journal of cell biology}, 190(2):165--175, 2010.

\bibitem{denk1990two}
Winfried Denk, James~H. Strickler, Watt~W. Webb, et~al.
\newblock Two-photon laser scanning fluorescence microscopy.
\newblock {\em Science}, 248(4951):73--76, 1990.

\bibitem{york2012resolution}
Andrew~G. York, Sapun~H. Parekh, Damian Dalle~Nogare, Robert~S. Fischer, Kelsey
  Temprine, Marina Mione, Ajay~B. Chitnis, Christian~A. Combs, and Hari Shroff.
\newblock Resolution doubling in live, multicellular organisms via multifocal
  structured illumination microscopy.
\newblock {\em Nature methods}, 9(7):749--754, 2012.

\bibitem{york2013instant}
Andrew~G. York, Panagiotis Chandris, Damian Dalle~Nogare, Jeffrey Head, Peter
  Wawrzusin, Robert~S. Fischer, Ajay Chitnis, and Hari Shroff.
\newblock Instant super-resolution imaging in live cells and embryos via analog
  image processing.
\newblock {\em Nature methods}, 10(11):1122--1126, 2013.

\bibitem{gerchberg1972practical}
Ralph~W. Gerchberg.
\newblock A practical algorithm for the determination of phase from image and
  diffraction plane pictures.
\newblock {\em Optik}, 35:237, 1972.

\bibitem{shroff2008otf}
Sapna~A. Shroff, James~R. Fienup, and David~R. Williams.
\newblock Otf compensation in structured illumination superresolution images.
\newblock In {\em Optical Engineering+ Applications}, pages 709402--709402.
  International Society for Optics and Photonics, 2008.

\bibitem{gustafsson2008three}
Mats~G.L. Gustafsson, Lin Shao, Peter~M. Carlton, C.J.~Rachel Wang, Inna~N.
  Golubovskaya, W.~Zacheus Cande, David~A. Agard, and John~W. Sedat.
\newblock Three-dimensional resolution doubling in wide-field fluorescence
  microscopy by structured illumination.
\newblock {\em Biophysical journal}, 94(12):4957--4970, 2008.

\bibitem{lal2016structured}
Amit Lal, Chunyan Shan, and Peng Xi.
\newblock Structured illumination microscopy image reconstruction algorithm.
\newblock {\em IEEE Journal of Selected Topics in Quantum Electronics},
  22(4):1--14, 2016.

\bibitem{kubby2013adaptive}
Joel~A. Kubby.
\newblock {\em Adaptive Optics for Biological Imaging}.
\newblock CRC press, 2013.

\bibitem{tao2011adaptive}
Xiaodong Tao, Bautista Fernandez, Oscar Azucena, Min Fu, Denise Garcia, Yi~Zuo,
  Diana~C. Chen, and Joel Kubby.
\newblock Adaptive optics confocal microscopy using direct wavefront sensing.
\newblock {\em Optics letters}, 36(7):1062--1064, 2011.

\bibitem{gu1996principles}
Min Gu.
\newblock {\em Principles of three dimensional imaging in confocal
  microscopes}.
\newblock World Scientific, 1996.

\bibitem{sheppard1990image}
C.J.R. Sheppard and Min Gu.
\newblock Image formation in two-photon fluorescence microscopy.
\newblock {\em Optik}, 86(3):104--106, 1990.

\end{thebibliography}

\end{document}